\documentclass[final]{ias2}

\usepackage{graphicx} 
\usepackage{multirow}
\usepackage{array} 

\usepackage{hyperref} 

\begin{document}

\markboth{A simplistic pedagogical formulation of a thermal speed distribution using a relativistic framework}{Ashmeet Singh}

\title{A simplistic pedagogical formulation of a thermal speed distribution using a relativistic framework}

\author[iitr]{Ashmeet Singh} 
\email{ashmtuph@iitr.ernet.in}
\address[iitr]{Department of Physics, Indian Institute of Technology Roorkee, Roorkee - 247667, Uttarakhand, India}

\begin{abstract}
A novel pedagogical technique is presented that can be used in the undergraduate (UG) class to formulate a relativistically extended Kinetic Theory of Gases and thermal speed distribution, while assuming the basic thermal symmetry arguments of the famous Maxwell-Boltzmann distribution as presented at the UG level. The adopted framework can be used by students to understand the physics in a thermally governed system at high temperature and speeds , without having to indulge in high level tensor based mathematics, as has been done by the previous works in the subject. Our approach, a logical extension of that proposed by Maxwell, will first recapitulate what is taught and known in the UG class and then present a methodology inspired from the Maxwell-Boltzmann framework that will help students to understand and derive the physics of relativistic thermal systems. The methodology uses simple tools well known to undergraduates and involves a component of computational techniques that can be used to involve students in this exercise. We have tried to place the current work in a larger perspective in regard to the earlier works done and emphasize on it's simplicity and accessibility to students. Towards the end, interesting implications of the relativistically extended distribution are presented and compared with the Maxwell-Boltzmann results at various temperatures. 
\end{abstract}


 
\maketitle

\section{Introduction}
\subsection{Thermal Gas in the Undergraduate Class}
The standard Kinetic Theory of Gases (KTG) as taught in a basic undergraduate (UG) course is a collision-based description of a classical ideal gas of particles (chemically homogeneous),\cite{classicalIdeal} treating them as rigid spheres which undergo perfect elastic collisions with the walls of the container using a Newtonian framework and assumptions.\cite{newtKTG} It results in expressions for the macroscopic properties of the gas (pressure, average kinetic energy {\it $\overline{K}$}, root mean square speed {\it $v_{\rm rms}$} etc.) as a function of the absolute temperature {\it T} of the gas. The derivation of these results can be found in commonly used texts,\cite{commonText} and yields the following major results for a ideal classical gas of {\it N} particles each having a rest mass {\it m} with {\it $k_{B}$} being the Boltzmann constant:
\begin{equation}
\label{eq:vrms2}
\overline{v^2} =\frac{3 k_{B} T}{m}	\: ,
\end{equation}
\begin{equation}
\label{eq:keavg}
\overline{K} =\frac{3N k_{B} T}{2}		\:	,
\end{equation}

where a bar over a quantity represents it's averaged value.  A distribution function describes the probability of a particle's speed near a particular value as a function of the gas' absolute temperature, the mass of the particle and the value of speed under consideration. Random thermal motion dictates that both position and velocity space are uniformly distributed and the distribution function is stationary with time.\cite{commonText} Symmetry between the three orthogonal components of velocity space has been asserted since no direction is preferred over the other in homogeneous space in a Newtonian framework, where each velocity component can take up any value \\ between $[0 , \infty )$. A probability distribution function(PDF) {\it f(.)} of the speed of the particles in one direction must also give the directional probability in other independent directions as well. In the cartesian coordinate system, the probability of the speed of a particle to lie between {\it $v_k \to v_k + dv_k$} for $k = \{x,y,z\}$ is given by {\it $f(v_x).f(v_y).f(v_z)dv_x dv_y dv_z$} as these individual probabilities are independent. Since any direction is as good as the others, the resultant distribution function {\it j(.)} must only depend on the total speed {\it $v$} of the particle. We shall assume factorability since it can be justified to a pedagogical extent that directional probabilities are interchangeable and independent, 
\begin{equation}
\label{eq:speedd}
j(v^2) = j(v_x^2 + v_y^2 + v_z^2) = f(v_x).f(v_y).f(v_z)	\: .
\end{equation}
The form of the directional function {\it f(.)} is clearly that of an exponential as seen by taking partial derivatives of eq.~(\ref{eq:speedd}) with respect to $v_k$ and relating it to the factorability argument which represents a Standard Probability Distribution function of the form,
\begin{equation}
\label{eq:MBdirectionalPDF}
f(v_k) = Ae^{-Bv_k^2}	\:	,
\end{equation}
  where {\it A} and {\it B} are constants depending on {\it T} and {\it m}. The Maxwell-Boltzmann Thermal speed distribution\cite{MBdistribution} (MB distribution) gives the probability {\it P($v$) dv} of the particle speed to lie within an elementary volume $dv_x$ $dv_y$ $dv_z$, or to lie between the speeds {\it $v$} and {\it $v + dv$} in the velocity space, centered on $(v_x,v_y,v_z)$ , where $dv_x$ $dv_y$ $dv_z$ $\equiv$ 4$\pi$ $v^2$ $dv$:
\begin{equation}
\label{eq:MBpdf}
P(v)dv = f(v_x).f(v_y).f(v_z)dv_x dv_y dv_z = 4\pi A^3 v^2 e^{-Bv^2} dv	\:	.
\end{equation}

The constants {\it A} and {\it B} in eq.~(\ref{eq:MBpdf}) are determined using the two integration conditions:
\begin{enumerate}
\item {\bf The Classical All Particle Condition}
\par
Classically and non-relativistically, the speed {\it $v$} of all particles follows 0 $\leq$ {\it $v$} $<$ $\infty$, hence the PDF {\it P($v$)} of eq.~(\ref{eq:MBpdf}) can be normalized as:
\begin{equation}
\label{eq:MBapc}
\int_0^\infty \! P(v) \, \mathrm{d} v = \int_0^\infty \! 4\pi A^3 v^2 e^{-Bv^2} \, \mathrm{d} v = 1	\:	.
\end{equation}
\item {\bf The Classical Kinetic Theory of Gases {\it $v_{rms}$} Result}
\par
Relating eq.~(\ref{eq:vrms2}) with the probability distribution function of eq.~(\ref{eq:MBpdf}), we get,
\begin{equation}
\label{eq:classicalKTG}
\overline{v^2} = \int_0^\infty \! v^2 P(v) \, \mathrm{d} v = \int_0^\infty \! 4\pi A^3 v^4 e^{-Bv^2} \, \mathrm{d} v = \frac{3 k_{B} T}{m}	\:	.
\end{equation}
\end{enumerate}
Thus,
\begin{equation}
\label{eq:maxwellAB}
A = {\left(\frac{m}{2\pi k_{B} T}\right)}^{1/2}       \quad  \quad   \quad   \quad   \quad   \quad   \quad                      B = {\left(\frac{m}{2 k_{B} T}\right)}	\:	.
\end{equation}
\subsection {Need and Motivation to extend the Maxwell-Boltzmann Distribution}
The following points on the formulation of the MB distribution yield discrepancies in regard to the Special Theory of Relativity:
\begin{enumerate}
\item The classical KTG is based on Newtonian mechanics and does not consider the constraints of Special Relativity. The Normal probability distribution that the MB distribution uses in eq.~(\ref{eq:MBpdf}) is a gaussian distribution and the variable, the speed of the particle {\it $v$}, extends till infinity. Classical MB distribution assumes that \\ {\it $v$} $\epsilon$ {[ 0, $\infty$ )} as in the normalization scheme of eq.~(\ref{eq:MBapc}), but Einstein's Special Theory of Relativity\cite{STR} limits {\it $v$} to {\it c}: the speed of light in free space. MB distribution at any {\it T} will always predict a non-zero fraction of particles to have speeds greater than {\it c}, which is physically forbidden and leads to inaccuracies in the distribution, particularly at high temperatures.
\item  As the particle gas approaches high temperatures, more particles will have {\it $v \to c$} and hence relativistic effects cannot be ignored.
    Thus, we cannot use $\overline{K} = ({m \overline{v^2}}/{2})$ as is implied by Eq~(\ref{eq:vrms2}). Rather, using Special Theory, $\overline{K} = ( \overline{\gamma} -1)mc^2$ where \\ $\gamma = {(1-{v^2}/{c^2})}^{-1/2}$ is the Lorentz factor. The basic result of the KTG in eq.~(\ref{eq:vrms2}) can be extended under the constraints of the Special Theory of Relativity as done in the next section.
\end{enumerate}
\begin{figure}[ht]
\includegraphics[width=1.0\columnwidth]{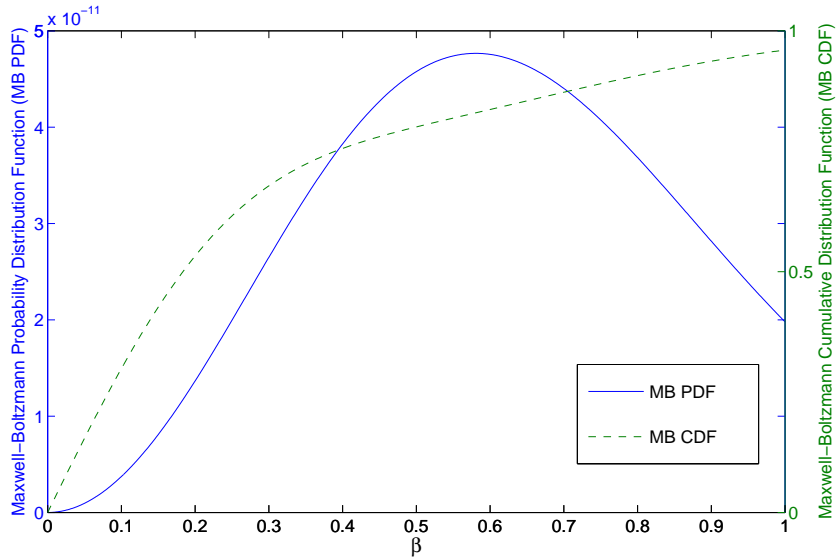}
\caption{MB PDF (Solid Curve) and CDF (Broken Curve) for a classical electron gas at T = $10^{9}$ K}
\label{10-9eg}
\end{figure} 
Astrophysical systems, e.g. the Intra cluster medium of galaxy clusters, can reach temperatures as high as $10^{8}\rm \: K$,\cite{ICM} which involve relativistic speeds and describing processes like Thermal Bremsstrahlung Emission\cite{bremsstrahlung} using MB distributions will lead to errors as depicted with the help of Fig.~\ref{10-9eg}. It depicts the MB distribution for a classical electron gas at $T = 10^{9}\rm \: K$ with $v_{\rm {most \: probable}} = 0.58 \:  c$ and shows that the Cumulative Distribution Function (CDF) $\neq$ 1 as $v \rightarrow c$ and predicts about 3.2 $\%$ particles to have speeds beyond the speed of light in vacuum. The error in the MB distribution is beyond negligence and is not an accurate description of the system. KTG and MB distributions are described and taught as elaborated in \S 1.1 and are used extensively in the UG curriculum to describe different thermal systems, whose speeds can be relativistic as shown above and to allow for deeper understanding of the system at student level, we are motivated to extend the classical KTG and MB distribution into the relativistic regime by incorporating the postulates of Special Theory of Relativity.
\section {Pedagogical extension of KTG into the relativistic regime}
\subsection{Adding Perspective}
In this paper, a novel approach to extend the KTG into the relativistic regime has been achieved, with elementary use of Four Vectors\cite{fourVector}, a standard tool of the undergraduate physics class. This approach allows one to understand the physics behind the macroscopic properties of a relativistic gas on lines similar to the standard KTG derivation using a collision based approach, but now under the constraints of Special Theory of Relativity. It is important to mention at this stage that the relativistic kinetic theory of gases is a very old and extremely well studied subject. It, hence, seems relevant to highlight a few key works in this regard. One of the most important and early relativistic extension of the KTG was done by J\"uttner in his two papers published in 1911\cite{juttner1911} and 1928\cite{juttner1928}, respectively. He derived the Maxwell-J\"uttner thermal speed distribution for relativistic gases which is a relativistic generalization of the celebrated Maxwell-Boltzmann thermal speed distribution. Following this, J. L. Synge in his book\cite{synge} presented a detailed analysis of a relativistic KTG using a four-dimensional geometric approach. Pioneering work by Israel\cite{israel1963} has developed the transport processes of a relativistic Boltzmann gas following the geometrical approach due to Synge. Israel describes these processes for a simple gas in a self-consistent gravitational field and close binary elastic collisions. One of his most important results is that a relativistic gas has a bulk viscosity which vanishes only in the classical limit. Many such interesting works have been largely possible due to the development of the relativistic Boltzmann Equation\cite{carlo_kramer_2002, groot_leeuwen_weert}. A consolidated and extensive account of the development and details of the KTG has been compiled in the series by Brush\cite{brush1957, brush1958, brush1961_1, brush1961_2}. Many of these past works in this regard are based on highly involved tensor-based calculus, a mathematical tool not available to all UG students at the time thermal systems are introduced to them. Our methodology, as presented in the next section, uses elementary Four Vectors and minimal mathematics to develop a relativistic KTG but tries to offer insight into the physics of the problem. Our extension gives the macroscopic properties of the relativistic gas including its average kinetic energy, most probable speed etc. A comparison with the methodologies of previous works has been presented in a later section.
\par
\subsection{The Relativistic KTG Derivation}
From an inertial frame of reference {\it S}, we observe a box of volume $V = l^3$ containing {\it N} identical classical particles with all standard KTG assumptions valid except that relativistic effects can be ignored. Let the $\rm {i^{th}}$ particle in the box be described by the position four vector $X$ $\equiv$ ($x^{0}$, $x$, $y$, $z$). Let $\Delta \tau$ be the proper time between events as measured in the particle's rest frame {\it{$S^{'}_{i}$}} and $\Delta$ $t$ be the improper time between events as measured in the frame {\it S} which are related by the Lorentz factor $\gamma$ as:

\begin{equation}
\label{eq:timedilation}
\frac{d\tau}{dt} = {(1-\frac{v^2}{c^2})}^{-1/2} =  \gamma	\:	.
\end{equation}
The corresponding acceleration four vector can be written in a form,\cite{fourAcc}
\begin{equation}
\label{eq:accfourvector}
A_{i} \equiv (\dot{\gamma} \gamma c, \gamma^{2} \vec{a} +  \dot{\gamma} \gamma \vec{v})	\:	,
\end{equation}
where $\vec{v}$  is the spatial velocity of the particle, $\vec{a}$ = ${d \vec{v}}/{dt}$ and $\dot{\gamma}$ = ${\gamma^{3} (\vec{a} \cdot \vec{v})}/{c^{2}}$. (A dot over a quantity represents it's time derivative.)

First, the particle-wall elastic collision in one dimension in the frame {\it S} is modeled and then the results are generalized to three dimensions. In a one dimensional analysis along an axis perpendicular to a particular wall of the container, say {\it x}-axis: $\vec{v} = v_{x} \: \hat{x}$ and the acceleration $\vec{a}$ = ${d \vec{v}}/{dt}$ = ${\Delta \vec{v}}/{\Delta t}$.
A similar approach as evolved in the Standard Newtonian KTG can be extended here using velocity and acceleration four vectors, instead of ``only-space" vectors, such that on each collision with the particular wall of the container, we have the following:
\begin{equation}
\Delta \vec{v} = 2 v_{x} \: \hat{x}	\:	,
\end{equation}
\begin{equation}
\Delta t = \frac{2l}{v_{x}}		\:	,
\end{equation}
\begin{equation}
\vec{a} = \frac{v^{2}}{l} \: \hat{x} = \frac{v^{2}_{x}}{l}  \: \hat{x}	\:	,
\end{equation}
\begin{equation}
\vec{a} \cdot \vec{v} = \frac{v^{3}_{x}}{l}	\:	.
\end{equation}
\par
Thus the spatial component of the acceleration four vector takes the form of,
\begin{equation}
A_{\rm {spatial,i,1D}} = \frac{({v^{2}})({{\gamma^{2}_{x}}})}{l} + \frac{({v^{4}})({{\gamma^{4}_{x}}})}{l {c^{2}}}  =  \frac{({v^{2}})({{\gamma^{2}_{x}}})}{l}\left[1 + \frac{({v^{2}})({{\gamma^{2}_{x}}})}{c^{2}}\right] \:	,
\end{equation}
where $\gamma_{x} = {(1-{v_{x}^2}/{c^2})}^{-1/2}$ is the Lorentz factor assigned in one dimension and \\ since $[1 + ({{v^{2}}({{\gamma^{2}_{x}}})}/{c^{2}})] = ({{\gamma^{2}_{x}}})$, hence we get:
\begin{equation}
A_{\rm {spatial,i,1D}} = \frac{({v^{2}_{x}})({{\gamma^{4}_{x}}})}{l}	\:	.
\end{equation}
The spatial part of the force four vector can be easily derived and shown in the last equation of this internet resource\cite{fourAcc} to be written as $F_{\rm {spatial,i,1D}} = m \: A_{\rm {spatial,i,1D}}$ and macroscopic gas properties can be derived as follows. The pressure due to the $i^{th}$ particle is
\begin{equation}
\label{eq:P-1D}
P_{\rm {i,1D}} = \frac{(m{v^{2}_{x}})({{\gamma^{4}_{x}}})}{V}	\:	.
\end{equation}
The total pressure of the gas can be found by summing $P_{i,1D}$ in eq.~(\ref{eq:P-1D}) over all the {\it N} particles in the box
\begin{equation}
\label{eq:Ptotal}
P_{\rm total} = \sum_{i=1}^N P_{i,1D} = \sum_{i=1}^N \frac{N m c^2}{V}  \:    \overline{({\beta^{2}_{x}})({{\gamma^{4}_{x}}})}	\:	,
\end{equation}
where $\beta = {v}/{c}$ and dimension-specific beta $\beta_{k} = {v_{k}}/{c}$ for $k = \{x,y,z\}$. \\
The particles of the gas are in constant random motion under thermal equilibrium which provides that all directions away from the walls of the container are identical in all physical respects and no particular direction is preferred over the other. The symmetry among all the three cartesian directions due to thermal equilibrium of the gas as described in \S 1.1 leads to the conclusion that the average contribution of each cartesian direction component is the same,
\begin{equation}
\label{eq:symmbeta}
\bar{\beta_x^2} = \bar{\beta_y^2} = \bar{\beta_z^2} = \frac{\bar{\beta^2}}{3}	\:	,
\end{equation} where {\it $\beta_x$}, {\it $\beta_y$}, {\it $\beta_z$} are related to the beta factor {\it $\beta$} of the particle as, {\it $\beta^2 = \beta_x^2 + \beta_y^2 + \beta_z^2$}. Mathematical analysis using infinite geometric series helps alternatively express eq.~(\ref{eq:Ptotal}) by introducing average of powers of $\beta$ using eq.~(\ref{eq:symmbeta}) as:
\begin{equation}
\label{eq:Ptotalfinal}
P_{\rm total} = \frac{3 N m c^2}{V}      \overline{({\beta^{2}})({{\theta^{4}}})}	\:	,
\end{equation}
where
\begin{equation}
\label{eq:theta2}
\theta^2 = \frac{1}{3 - \beta^2}	\:	.
\end{equation} \\
As with the classical KTG, we assume that the gas particles behave ideally and follow the Ideal Gas Equation ($P V = N k_{B} T$) to generate the final Relativistic KTG postulate
\begin{equation}
\label{eq:relKTG}
\overline{({\beta^{2}})({{\theta^{4}}})} = \frac{k_{B} T}{3 m c^2}	\:	.
\end{equation}
Equation~(\ref{eq:relKTG}) is the relativistically extended version of the classical KTG postulate as in eq.~(\ref{eq:vrms2}) by incorporating constraints of Special Relativity and can supersede the classical KTG postulate.

\section {Formulating a Relativistic Distribution Function}
The underlying pedagogical symmetry arguments are the defining criteria for random motion in thermal equilibrium in the MB distribution. The uniform occupation of position and velocity space and equivalence among different directions as stated in \S  1.1 are assumed to be valid. This can be thought so because any triplet of velocity components $(v_{i}, v_{j}, v_{k})$ can as well be valid by interchanging the components between any two directions of velocity space, if none has fixed values.

It is worth mentioning at this stage that the derivation of the MB distribution presented in \S 1.1 using statistical independence of velocity components is considered invalid in a strict sense.\cite{dunbar} This approach is commonly used in pedagogical literature to emphasize on the key aspects of the distribution. The correct derivation of the MB thermal distribution can be found in the book by Richet.\cite{richet} We can formulate the relativistically extended (RE) distribution function by incorporating changes in eq.~(\ref{eq:MBpdf}) compliant with Special Relativity. Let $J(v)dv$ be the RE distribution function that has all the features of a thermal distribution. Postulates of Special Relativity will dictate that $J(0) = J(v \geq c) = 0$ because no particle can travel with the vacuum speed of light or beyond. The following inequality must hold,
\begin{equation}
\label{vlessthanc}
({v_{x}}^{2} \: + \: {v_{y}}^{2} \: + \: {v_{z}}^{2} \: = \: v^2 \:) < \: c^2
\end{equation}
This implies that the possibilities of taking up values in a triplet $(v_{x}, v_{y}, v_{z})$ depends on the total speed $v$ of the particle and also the values fixed for individual components $v_{i}$ for $i = {x,y,z}$. This argument backed by the statistical interchangeability of each velocity component asserts that the $i^{\rm {th}}$ directional PDF depends not only on $v_{i}$ but also on $v$ and the dependence being the same for all directions.

The factorability of eq.~(\ref{eq:speedd}) will hold as emphasized above for the functional form of $J(v)dv$ and the directional distribution function $g_{\rm RE}(.)$ can be taken to be \\ $g_{\rm RE}(v_{k}) = f_{\rm MB}(\alpha (v) \: v_{k})$
 where $f_{MB}$ is the MB directional PDF of eq.~(\ref{eq:MBdirectionalPDF}) and $P$ and $R$ are constants. In this equation, $\alpha (v)$ is the total speed dependence in the distribution functions as discussed earlier in this section. It must hold that $\alpha (c) \rightarrow \infty$ so that $g_{\rm RE} \rightarrow 0$ as any velocity component approaches the speed of light.
  \par
 Consider the following excerpt of calculation,
 \begin{equation}
 \label{alpha1}
 \gamma^{2} \: = \: \frac{1}{1 - (\frac{v^2}{c^2})}
 \end{equation}
  \begin{equation}
 \label{alpha2}
\frac{v^2}{c^2} \: = \: 1 - \frac{1}{\gamma^2}
 \end{equation}
 or equivalently,
  \begin{equation}
 \label{eq:alpha3}
 (\gamma v)^{2} \: = \: c^{2} (\gamma^{2} - 1)
 \end{equation}
 Using the form of eq.~(\ref{eq:alpha3}), we can define the function $\alpha (v) \: \equiv \: \gamma$ which has all the required properties and RE directional PDF can be obtained by replacing $v_{k}$ in the MB directional PDF by $\alpha(v) v_{k}$ as follows,
\begin{equation}
\label{eq:REdirectionalPDF}
g_{\rm RE}(v_{k}) = P \: {\rm exp}{(-R v_k^2 \gamma^2)}
\end{equation}
Thus, we now have a distribution function that depends on the total speed $v$ of the particle, abides by the constraints of STR and yet has features of thermal independence of velocity components. The resultant RE distribution function in velocity space centered on $(v_{x}, v_{y}, v_{z})$ or equivalently in $\beta$ space centered on $(\beta_{x}, \beta_{y}, \beta_{z})$ is, respectively,

\begin{equation}
\label{eq:REprf2}
J(v) dv = 4 \pi \: P^3 \: v^2 {\rm exp}{(-R v^2 \gamma^2)} dv	\:	,
\end{equation}
\begin{equation}
\label{eq:REprf}
J(\beta) d \beta = 4 \pi \: X^3 \: \beta^2 {\rm exp}{(-Y \beta^2 \gamma^2)} d \beta	\:	, 
\end{equation}
and the constants in these equations being related as $X = P c$ and $Y = R c^2$

\begin{figure}[h!]
\begin{center}
\includegraphics[width=1.0\columnwidth]{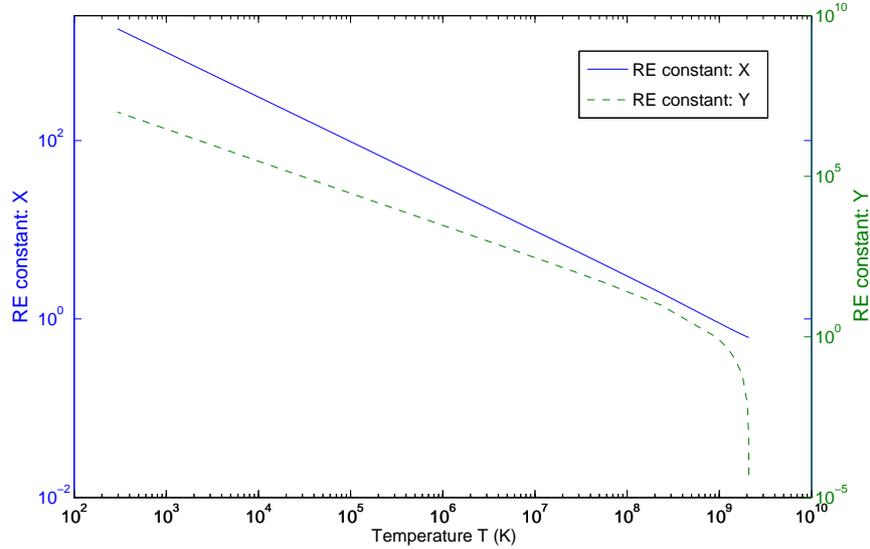}
\caption{Temperature dependance of constants X (Solid Curve) and Y (Broken Curve) of eq.~(\ref{eq:REprf})}
\label{constantsPlot}
\end{center}
\end{figure}

It is important to mention at this stage that $J(\beta) d \beta$ is not strictly a Normal Probability Distribution\cite{normalDistribution} because the argument of the exponential in eq.~(\ref{eq:REprf}) does not vary as $\beta^2$ but as $\beta^2 \gamma^2$ i.e. $\left[\beta^2/{(1 - \beta^2)}\right]$. Hence, eq.~(\ref{eq:REprf}) is a Probability Representative Function(PRF) and not strictly a PDF since it predicts a number proportional to the probability.
The constants {\it X} and {\it Y} in eq.~(\ref{eq:REprf}) are dependant on {\it T} and {\it m}, and can be determined using the following two conditions:
\begin{enumerate}
\item {\bf The Relativistic All Particle Condition}
\par
As constrained by the Special Theory of Relativity, the speed {\it ${\it v}$} of any particle follows \\ 0 $\leq$ {\it $v$} $<$ $c$, hence the PRF  $J(\beta) d \beta$ of eq.~(\ref{eq:REprf}) can be normalized to include all the {\it N} gas particles as:
\begin{equation}
\label{eq:REapc}
\int_0^1 \! J(\beta) \, \mathrm{d} \beta = \int_0^1 \! 4\pi X^3 \beta^2 e^{-Y \beta^2 \gamma^2} \, \mathrm{d} \beta = 1	\:	.
\end{equation}
\item {\bf The Relativistically Extended KTG Result}
\par
Relating eq.~(\ref{eq:relKTG}) and ~(\ref{eq:theta2}) with the PRF of eq.~(\ref{eq:REprf}), one gets:
\begin{equation}
\label{eq:reKTG}
\overline{\beta^2 \theta^4} = \int_0^1 \! \beta^2 \theta^4 J(\beta) \, \mathrm{d} \beta = \int_0^1 \! 4\pi X^3 \frac{\beta^4}{(3 - \beta^2)^2} e^{-Y \beta^2 \gamma^2} \, \mathrm{d} \beta = \frac{k_{B} T}{3 m c^2}	\:	.
\end{equation}
\end{enumerate}
The integral conditions of eq.~(\ref{eq:REapc}) and ~(\ref{eq:reKTG}) are not solvable in closed analytic form. The computational facility Wolfram Alpha\cite{wolfram} available  was used to numerically solve these for a classical ideal gas of electrons by feeding a 40-value long array of constant {\it Y} to generate values of corresponding {\it X} and {\it T.} The Curve Fitting Tool of MATLAB\cite{matlab} can be used to plot the temperature dependance of {\it X} and {\it Y} (Fig.~\ref{constantsPlot}) and apply a best fit for the data wherever possible.
\\A linear polynomial best fit the correlation of ln(X) - ln(T) and the {\it T} dependance of {\it X} is
\begin{equation}
\label{eq:lnX-lnT}
\rm{ln} \: X = 10.4 - ( 0.5063 )\: \rm {ln} \: T	\:	.
\end{equation}

\section{Features of the Relativistically Extended Distribution}
\subsection{The Most Probable Speed Analysis}
The most probable (mp) speed of the particles in the MB distribution is $v_{\rm mp} = {({2 k_{B} T}/{m})}^{1/2}$. A similar expression using the RE distribution can be found when the PRF of eq.~(\ref{eq:REprf}) is maximum, which gives
\begin{equation}
\label{eq:REmp}
\frac{\beta^2_{\rm mp}}{(1 - \beta^2_{\rm mp})^2} = \frac{1}{Y}		\:	,
\end{equation}
which can be solved using a quadratic analysis of roots. In the non-relativistic limit \\ $v << c$ so that $\gamma \rightarrow 1$, $\beta^2_{\rm mp}$ $\approx$ $\left[({1}/{Y}) = ({1}/{R c^2})\right]$ from eq.~(\ref{eq:REprf}) and ~(\ref{eq:REprf2}), which is equivalent to   $v^2_{\rm mp} \approx ({1}/{R})$. The constant {\it R} in the MB PDF of eq.~(\ref{eq:MBpdf}) is $\left( \frac{m}{2 k_{B} T} \right)$ which gives us,
\begin{equation}
\label{compareMP}
v_{\rm {mp, v<<c}} = {(\frac{2 k_{B} T}{m})}^{1/2} \:	,
\end{equation}
and hence in the non-relativistic regime, the results for most probable speed using the RE distribution match the MB results.
\subsection{The nature of temperature dependance of constant {\it Y}}
The argument of the exponential in the PRF of eq.~(\ref{eq:REprf2}) must be unit-less, hence {\it R} will have the units of $(\rm {{sec^2}\:{m^{-2}}})$ or equivalently $(\rm {{kg}\:{joule^{-1}}})$ i.e inverse of specific work. {\it R} in eq.~(\ref{eq:REprf2}) is inversely proportional to the work or energy required to heat the gas to the temperature {\it T}. The temperature dependance of {\it Y} as in Fig.~\ref{constantsPlot} is analogous to the \\ x-axis reflected plot of total energy $E = ({m c^2}/{\sqrt{1 - \beta^2}})$ of a particle with its $\beta$ parameter. As the temperature and the required energy of the gas to accomplish this increases there is decrease in {\it Y}. After a certain {\it T} range, any further increase in {\it T} requires an uncontrollable increase in energy input and a corresponding uncontrollable decrease in {\it Y}. It is clear that a high {\it T} involves relativistic speeds and nature demands lot of energy to sustain such high speeds and temperatures.

\section{A Comparison Study with the MB Distribution}
We now compare results of the MB and RE distributions at various temperatures and show that results of MB distribution for $T > 5 \times 10^8 \rm {K}$ are inaccurate beyond negligence and the RE distribution results are physically convincing and can be a good approximation to the real system.
As stated in \S 3, eq.~(\ref{eq:REprf}) is a PRF and predicts a number proportional to the probability. The MB and RE distributions can be compared at a given {\it T} by a parameter {\it (r)} at various speeds from eq.~(\ref{eq:MBpdf}) and eq.~(\ref{eq:REprf2}).
\begin{equation}
\label{eq:rparameter}
{\rm {Ratio}} (r) = \frac{P(v)}{J(v)}	\:	.
\end{equation}
We will use $r_{\rm max}$ to represent the Ratio {\it (r)} at the highest plotted speed and $r_{\rm min}$ will represent the Ratio {\it (r)} at the lowest plotted speed and $\Delta r = r_{\rm max} - r_{\rm min}$ , an average estimate of the scatter between the two functions at a given {\it T}.\\
\begin{figure}[h!]
\begin{center}
\includegraphics[width=1.0\columnwidth]{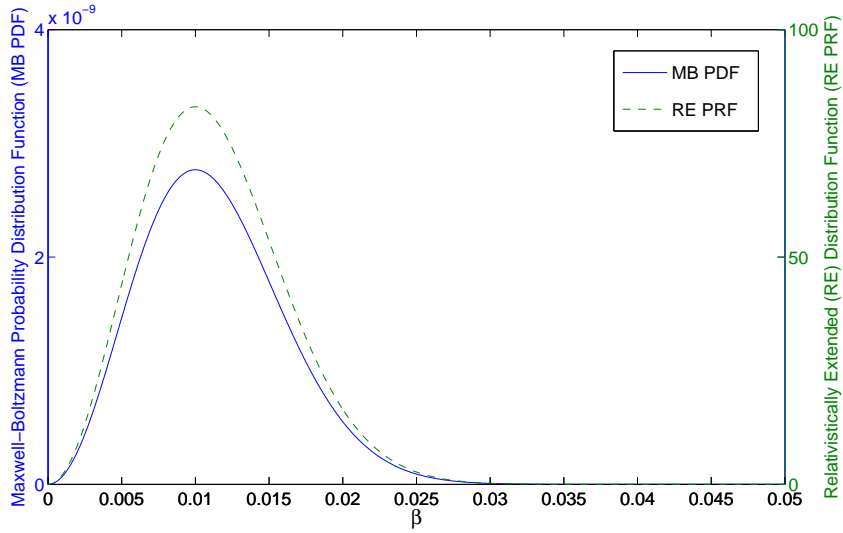}
\caption{MB PDF (Solid Curve) and Relativistically Extended (RE) PRF (Broken Curve) for T = 2.96 $\times 10^ 5$ K}
\label{medT}
\end{center}
\end{figure}
\begin{figure}[h!]
\begin{center}
\includegraphics[width=1.0\columnwidth]{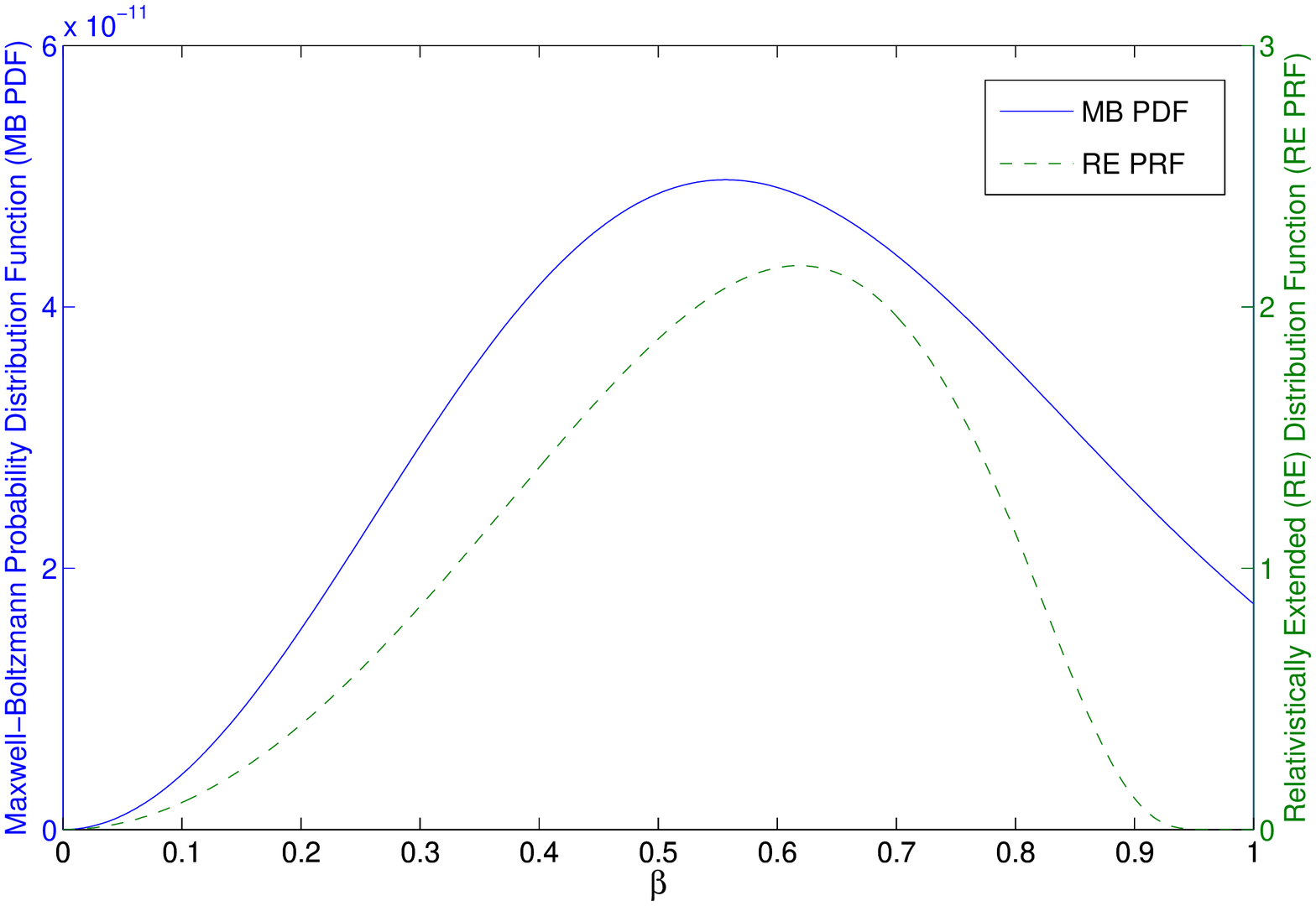}
\caption{MB PDF (Solid Curve) and Relativistically Extended (RE) PRF (Broken Curve) for T = 9.18 $\times 10^{8}$ K}
\label{highT}
\end{center}
\end{figure}
\begin{table}[h!]
\begin{center}
\caption{Comparing MB and RE distribution functions at two different temperatures}
\begin{tabular}{|c||c|} \hline
T = $2.96 \times 10^{5}$ K & T = $9.18 \times 10^{8}$ K \\ [1ex]
\hline
\hline
$r_{\rm max} = 3.52\times10^{-11}$ & $r_{\rm max} = 4.31\times10^{9}$\\
\hline
$r_{\rm min} = 3.53\times10^{-11}$ & $r_{\rm min} = 4.16\times10^{-11}$\\
\hline
$\Delta r = 1.86\times10^{-12}$ & $\Delta r = 4.31\times10^{9}$\\
\hline
$v_{\rm (mp - MB)} = 0.0099984 c$ & $v_{\rm (mp - MB)} = 0.5565418 c$ \\
\hline
$v_{\rm (mp - RE)} = 0.009999 c$ & $v_{\rm (mp - RE)} = 0.618034 c$ \\
\hline
$100\times \int_c^\infty \! P(v) \, \mathrm{d} v = 0 $ & $100\times \int_c^\infty \! P(v) \, \mathrm{d} v = 3.2\%  $ \\
\hline
$\int_c^\infty \! J(v) \, \mathrm{d} v = 0 $ & $\int_c^\infty \! J(v) \, \mathrm{d} v = 0 $ \\
\hline
\end{tabular}
\end{center}
\end{table}
Table I chalks out a comparison (Fig.~\ref{medT} and~\ref{highT}) between the MB and RE distribution functions at two different temperatures by comparing values of various parameters, representative of each distribution. As evident from the comparison table, for $T < 5 \times 10^8 \rm {\: K}$, MB and RE distributions match very well with the shape of the distribution function preserved and and the scatter parameter {\it r} between the two is of the order of $10^{-12}$ or less. RE distribution converges to Maxwellian results at low speeds and low temperatures.

Figures~\ref{medT} and~\ref{highT}  plot MB and RE distribution functions at the two comparison temperatures. They clearly show that for higher {\it T}, the shape of the two functions varies significantly, with MB predicting about 3.2$\%$ particles beyond {\it c}, while the RE function is a good description of the system under the constraints of special relativity.
\begin{figure}[h!]
\begin{center}
\includegraphics[width=1.0\columnwidth]{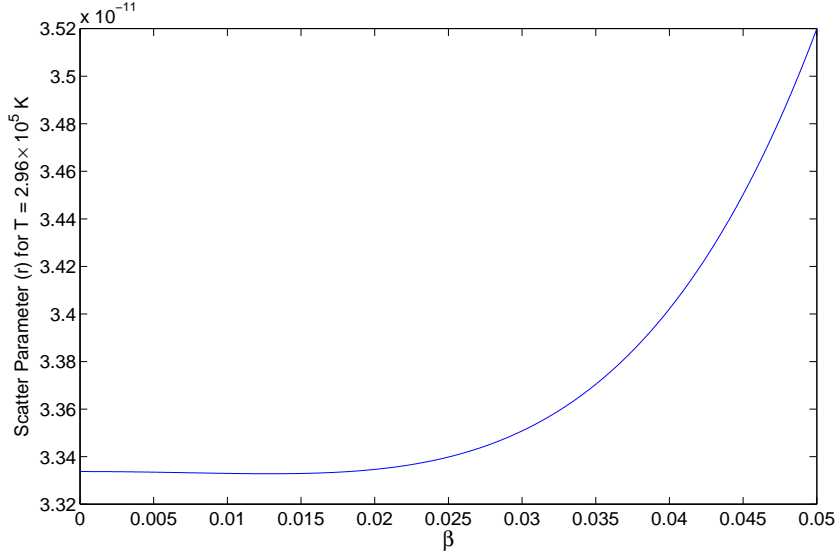}
\caption{Scatter Ratio "r" for T = $2.96 \times 10^{5}$ K}
\label{rScatter}
\end{center}
\end{figure}
Figure~\ref{rScatter} shows the variation of the Ratio {\it r} with $\beta$ for $T = 2.96 \times 10^{5}$ K.
 The scatter in the two distributions is extremely large as $v \rightarrow c$ but it is important to notice that at all temperature regimes, RE results converge to MB at low speeds. Overall, the RE and MB distributions diverge at high {\it T} and high {\it $v$} and the RE distribution favors higher speeds compared to MB, as shown by the $v_{\rm mp}$.

\section{Comparison with previous works on Relativistic KTG}
As has been pointed out earlier, the relativistic kinetic theory of gases is an old and well studied subject for more than a 100 years now. Previous works have been successful in describing the macroscopic properties of a relativistic gas, its distribution function and transport coefficients and other bulk properties, to mention a few. These have utilized various approaches to the problem, some geometric in nature while others completely based on an analytical treatment of the relativistic Boltzmann equation. The mathematical and physical framework, that has been used, is undoubtedly strong and sound but this level of sophistication is not usually available with UG students. Our methodology, in this paper, has been to focus on the broader physical aspects of a relativistic KTG. We have adopted a mathematically simple path that can make this work interesting and reachable to students. In such a formulation, some of the deeper subtleties of the subject are felt to be lost that have been covered in a more involved and complete way by previous works. We have been able to successfully derive the basic KTG and the macroscopic properties of the gas, but it is insufficient to understand the transport phenomenon and other bulk properties of the gas like viscosity and thermal conductivity. In addition to this, our method only handles non-interacting classical gas particles in complete thermal equilibrium, neglecting the quantum nature of the particles. Earlier works have focused on the problem of a collisional gas in various conditions deviating from equilibrium. Lee\cite{lee1950} has estimated the thermal conductivity of a relativistic Fermi gas, hence bringing in the quantum nature of the gas particles. Our approach tries to involve the reader by presenting simple, yet intuitive arguments based on the Special Theory of Relativity and a particle-wall collision-based KTG. The distribution function has been argumentatively extrapolated and seems to well describe the particle speeds. This, however does not reproduce the results of \cite{juttner1911} and \cite{synge} but contains the important physical aspects at the UG level. We feel that this work can be used to introduce students at the UG level to relativistic thermal systems to help them understand the basics. Interested students can further be referred to earlier works that will supplement their understanding.  

\section{Conclusion}
In our analysis of including Special Relativity in the KTG and a thermal distribution, we have presented a pedagogically useful method to introduce relativistic constraints that preserve the basic collision and thermal framework that is taught as the standard approach to this subject. The method of Four Vectors and particle-wall collisions along with the usage of computational techniques to solve integral conditions can be very effectively used in the UG classroom paradigm to help students appreciate the interlink between different areas of Physics and themselves try and formulate it. The ideas presented here give a reasonable description of the physical system but it is important to note that this approach is based on incorporation of Special Relativity in a previously accepted methodology and the formulation of the RE function involves assuming a logically sound function, extrapolating from the MB framework. This is a well studied subject with a lot of details and results been published earlier. Our work does not reproduce these results, both in formulation and extent but still presents a simple and intuitive approach for the students to study relativistic KTG. Thus, we suggest that due care must be taken while adopting our approach, which we feel is a good pedagogical tool in the UG classroom.
\begin{acknowledgments}
I am most grateful to Prof. Joydeep Bagchi and Dr. Surajit Paul of the Inter University Center for Astronomy and Astrophysics (IUCAA), Pune, India for their support, guidance and motivation. I am also grateful to Prof. Tashi Nautiyal of the Indian Institute of Technology Roorkee, Roorkee, India for kindly consenting to proofread the manuscript. 
\end{acknowledgments}


\end{document}